# Correlation of action potentials in adjacent neurons

## M. N. Shneider [1] and M. Pekker [2]

[1]Department of Mechanical and Aerospace Engineering, Princeton University, Princeton, NJ 08544, USA
[2]MMSolution, 6808 Walker Str., Philadelphia, PA 19135, USA

E-mail: m.n.shneider@gmail.com and pekkerm@gmail.com

**Abstract**

A possible mechanism for the synchronization of action potential propagation along a bundle of neurons (ephaptic coupling) is considered. It is shown that this mechanism is similar to the salutatory conduction of the action potential between the nodes of Ranvier in myelinated axons. The proposed model allows us to estimate the scale of the correlation, i.e., the distance between neurons in the nervous tissue, wherein their synchronization becomes possible. The possibility for experimental verification of the proposed model of synchronization is discussed.

## I. Introduction

In a recent paper [1] the appearance of synchronization spikes in four neurons located within 100 μm (the distance between initial segments did not exceed 10-15 microns) was observed. In this work it was suggested that a mechanism responsible for the observed synchronization of neurons may be the change of the electric potential near neural somas (initial segment) caused by spikes in one of the neurons. Generally speaking, this kind of interaction between neurons has been known for a long time and is called ephaptic coupling. The study of the ephaptic coupling effect caused by the currents in the pericellular space, induced by the currents through the excited membrane of axons, began with the works [2,3] (a contemporary detailed overview of the ephaptic coupling problem, can be found in [4]). Among the large series of works devoted to the synchronization of action potentials in the nearby neurons ref. [5,6], should be noted, in which the problem of interaction between the two non-myelined axons were considered on the basis of the Hodgkin-Huxley model. In these papers interactions of axons were described by the introduction of cross-terms, which were determined by the product of the current in a parallel axon and the corresponding coupling coefficient. However, based on the results of [5,6] and other similar works, it is difficult to estimate the characteristic scale of the correlation between two neurons, since the description of the relationship between axons in these studies is essentially phenomenological.

It is well known (see, eg [7]), that to excite an action potential in the nerve fiber it is necessary to apply a threshold voltage of 10-20 mV for a duration ~ 0.1 ms to the membrane of the non-myelined area of axon. The threshold conditions for the excitation of the action potential depends on many factors: temperature, local surface density of transmembrane sodium channel, etc [7-9].

From [10,11], in which the magnetic field generated by currents of the action potential were studied, it follows that propagating action potential causes a change in the potential on the outer surface of the membrane of just a few microvolts. We emphasize that it is not a potential difference between the outer and inner surfaces of the membrane but, namely, the potential on the outer surface of the membrane. It would not seem that such a small change in the surface potential of the membrane of the axon would excite an action potential in neighboring neurons. For this reason, the interaction of neurons was considered impossible for most mammalian nerve tissues [12-14].

In this paper the synchronization mechanism between nearby neurons (ephaptic coupling) is considered (Figure 1): based on the fact that the propagating action potential along an axon is always accompanied by currents in the physiological saline in the vicinity of the membrane. These currents may cause charging of the axons membranes of neighboring neurons, leading to the appearance of potential difference across the membrane, which is greater than the threshold of the action potential initiation. The estimations of the scale of the correlations between the parallel initial segments of the myelin neurons, obtained in this paper, are in agreement with the experimental data [1]. It is shown that the ephaptic coupling mechanism, to some extent, is similar to saltatory conduction of the action potential between the nodes of Ranvier in myelinated axons [4,15-18]. Such an approach allows obtaining estimates of the synchronization area (scale of correlation) of action potentials both for the case of myelin fibers as well as non-myelin.

The second part of this paper considers the longitudinal transfer of the potential in the myelin fiber. Estimations of the maximum longitudinal size of myelinated sections of the axon are obtained, when the propagation of the action potential is possible. In Part 3 a synchronization mechanism of neighboring neurons is considered. In part 4, the correlation region for the initiation of the action potentials in initial segments of myelinated axons is computed, using the formulas derived in part 3. In Part 5 the action potential excitation in the myelined axon by the action potential propagating along the neighboring myelined axon (Figure 1) is calculated. In Part 6, the conditions for the synchronization of action potentials of squid axons are obtained, which can be used to test the proposed mechanism for the synchronization of neurons.

## II. Saltatory conduction of the action potential in the myelin fiber.

Neurons in many vertebrates are surrounded by an electrically insulating myelin sheath, which provides a saltatory conduction of the action potential, serving to speed up the transmission of nerve impulses.[15-18] The myelin sheaths produced by different cells are separated by gaps, known as the nodes of Ranvier. These nodes support a saltatory conduction of action potentials, facilitating rapid propagation of nerve signals

In the process of the saltatory conduction of the action potential, the propagating impulse jumps from one Ranvier's node to the next. In each node of Ranvier (non-myelin section), when the potential difference across the membrane exceeds the threshold, an action potential is initiated. This action potential charges the next myelinated section, which is, in a certain sense, a long line with the leakage. The length of the myelined section is limited by the condition that the potential difference during a certain time exceed the threshold voltage sufficient to initiate an action potential at the next node of Ranvier. Saltatory conduction of the action potential in the myelin fiber is well studied (see for example, [16-19]).

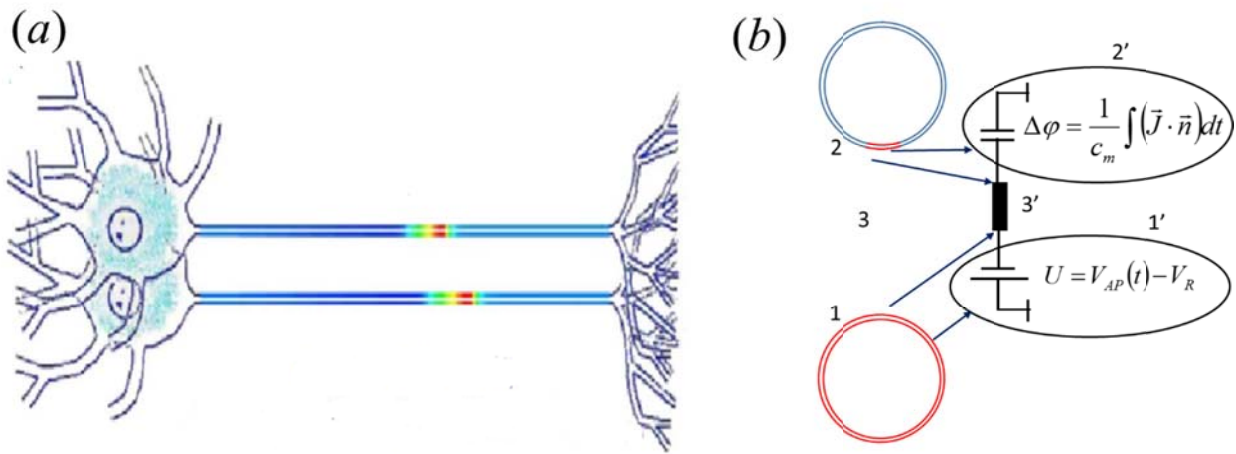

**Fig. 1.** Scheme illustrating interaction between neighbouring neurons. ($a$) – propagation of the action potential along the two correlated neurons. ($b$) – excitation of the action potential in the inactive axon by the active axon. 1 – active axon; 2 – inactive axon; in red shows a portion of the membrane being charged by currents, induced in saline 3; 1', 2', 3' – an equivalent circuit diagram of the charging surface of the membrane of the inactive axon. $c_m$ – capacity of the axon membrane per unit area, $\vec{J}$ - local charging current of the inactive axon membrane, $\vec{n}$ – normal unit vector to the surface of the membrane, $\Delta\varphi$ – local potential difference between the capacitor plates (outer and inner surfaces of the membrane), $U$ – EMF source (potential difference between the surfaces of the membrane of the axon, along which the action potential propagates), $V_{AP}$ – action potential, $V_R$ – the resting potential of the axon.

To examine the action potentials on myelinated nerve fibers, we employ the Goldman–Albus model of a neuron of a toad [17]. In this model, which is reproduced in this section for the convenience of the readers, a myelinated nerve fiber is represented as a leaky transmission line with uniform internode sections with a length $L_M$ and an outer diameter $D_M$ (myelin sheath inclusive) separated by nodes of Ranvier with a length $L_R$ and outer diameter $d_R$, which is equal to the diameter of an axon inside the myelin coating (Fig. 2).

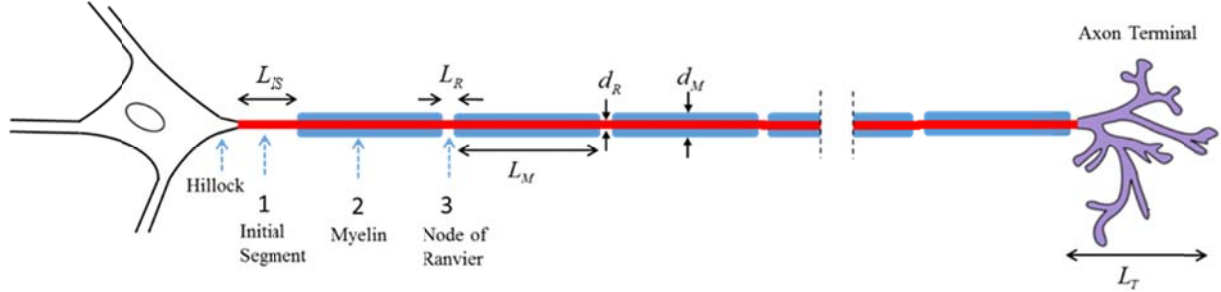

**Fig.2.** Scheme of a neuron with the characteristic elements of the myelinated axon.

Typically the size of $L_{IS}$ is in the range 20-50 μm, but it can reach 75 μm and even, 200 μm, $L_R \approx 1 - 2.5\,\mu m, L_M \sim 100 - 2000 \mu m, \ L_T \sim 10 - 2000$ μm [7].

The voltage $U(x,t)$ across the membrane in the internode region is found as a solution to the equation [17]:

$$\frac{\partial U}{\partial t} = a\frac{\partial^2 U}{\partial z^2} + bU , \qquad\qquad (1)$$

Where $a = 1/R_1 C_1; \ \ b = -1/R_m C_1$. Здесь $C_1 = k_1 / \ln(d_M / d_R)$, $R_1 = \dfrac{R_i}{\pi (d/2)^2}$, are the myelin capacitance and resistance per unit length; $R_m = k_2 / \ln(d_M / d_R)$ is the myelin resistance times unit length, and $k_1, k_2$ are constants; $R_i$ is the axoplasm specific resistance.

Equation (1) is the equation of the potential diffusion. Assume that at x = 0 is a node of Ranvier in which the action potential is initiated. Taking into account that $L_M \gg L_R$, the propagation of the potential diffusion wave along the myelined section can be considered on a semiaxis $0 \le z < \infty$ with the initial and boundary conditions:

$$\begin{aligned} U(z \ne 0, t = 0) &= 0 \\ U(z = 0, t) &= V_{AP}(t) - V_R \end{aligned} \qquad\qquad (2)$$

Here $V_{AP}(t)$ is the action potential in the Ranvier node, which is defined by the Frankenhaeuser–Huxley equation [19]; $V_R = -70$ mV is the corresponding resting potential. Using the coefficients and parameters given in [17,20] (we consider as an example, the myelined toad's axon), the action potential in the nodes of Ranvier has the form shown in Fig. 3. Fig. 4 shows the distributions the action potential at different times [20] calculated in the framework of approximations and the data [17].

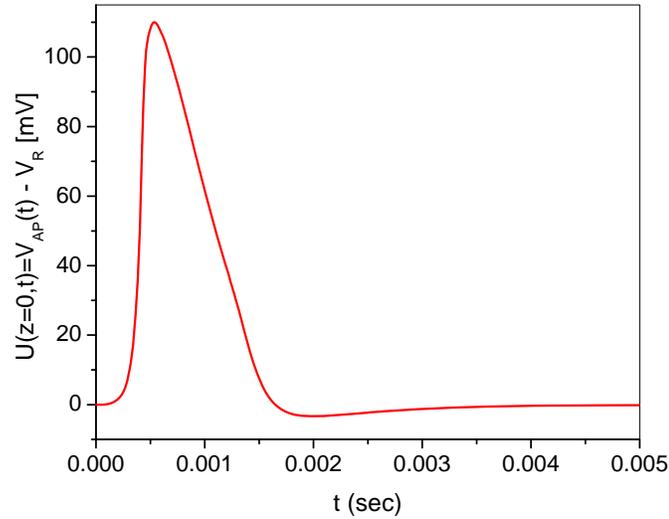

**Fig. 3.** The time dependence of the action potential at the nodes of Ranvier of a toad axon.

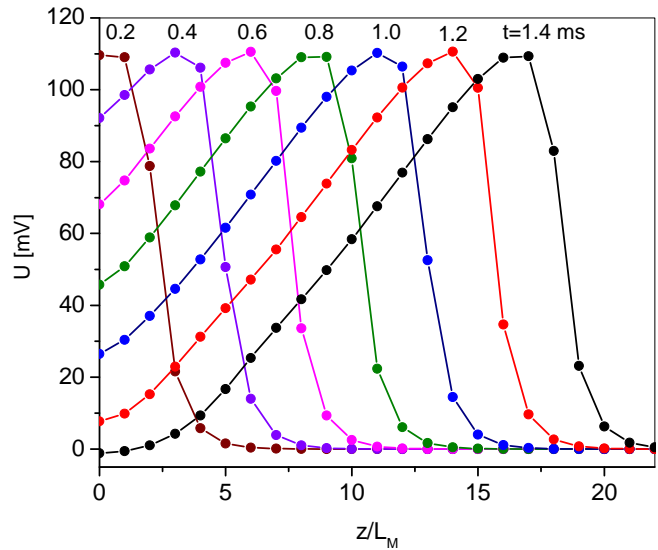

**Fig. 4.** The spatial profiles of the action potential. The nodes of Ranvier located (shown by circles), at $z/L_M$ =1,2,3… at the different instants of time.

Equation (1), with the conditions (2), represents the first boundary value problem of the diffusion equation with the general solution [21,22]

$$U(z,t) = \int\limits_0^t V_{AP}(\tau) H(t-\tau, z) d\tau$$

$$H(z,t) = \frac{z}{2\sqrt{\pi a} t^{3/2}} \cdot \exp\left(bt - \frac{z^2}{4at}\right),$$

(3)

Turning to the variable $\mu = 0.5z/\sqrt{a(t-\tau)}$, equation (3) is converted to a form suitable for the estimations

$$U(z,t) = \frac{2}{\sqrt{\pi}} \int\limits_{0.5z/\sqrt{at}}^{\infty} V_{AP}\left(t - \frac{z^2}{4a\mu^2}\right) \exp\left(\frac{bz^2}{4a\mu^2} - \mu^2\right) d\mu .$$

(4)

As already mentioned, the solution by the formula (4) for the evolution of the potential difference distribution along the myelined part was carried out for the parameters of the axon corresponding to a toad sciatic nerve, given in [17,20]: $d_M = 15\,\mu\text{m}$, $d_R = 9\,\mu\text{m}$; $k_1 = 16\ln1.43\,[\text{pF/cm}]$; $k_2 = 2.9 \cdot 10^7 / \ln 1.43\,[\Omega \cdot \text{cm}]$ ; $R_i = 110\,[\Omega \cdot \text{cm}]$.

A sample of calculation for these parameters is shown in Fig. 5. Assuming the threshold potential difference across the membrane $\Delta V \approx 20\,\text{mV}$ and the characteristic time interval required for the initiation of an action potential, $\Delta t \approx 0.1$ ms, we find from Fig. 5, that the length of myelined section cannot be longer than $L_M \approx 5$ mm. Note that this estimate of the limiting length is more than twice the length of a typical myelined segment of the toad's axon ($\sim 2$ mm). This is likely due to the fact that nature selects the length of myelined segment with a large margin, to ensure certain saltatory conduction of the action potential along the axon, even at significant variations in the parameters of the nerve fiber.

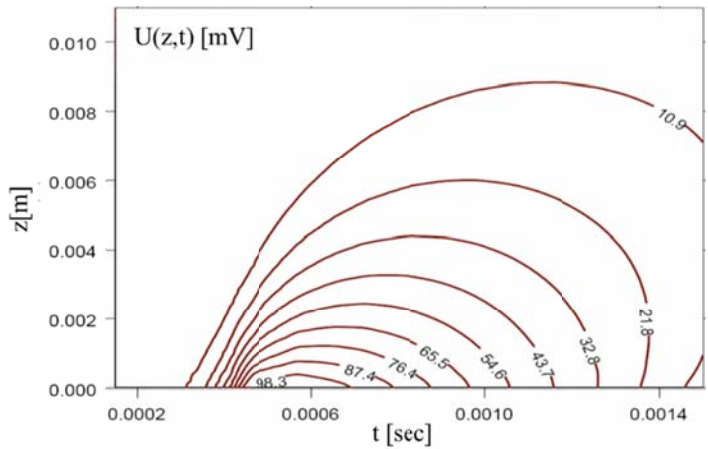

**Fig. 5.** Spatial-temporal dynamics of the myelin segment charging by the action potential in the node of Ranvier.

**III. The dynamics of potential and currents in the vicinity of a neuron in the case of the propagating action potentials**

Saline, the fluid where the neurons are located in living organisms is a highly conductive electrolyte, with $\sigma \sim 1-3 \; \Omega^{-1}\mathrm{m}^{-1}$ [23]. Therefore, the relaxation time of the volume charge in it (the Maxwell time) is of order $\tau_M = \varepsilon\varepsilon_0 / \sigma \sim 10^{-9}\mathrm{s}$ [24], where $\varepsilon_0$ is the dielectric constant of the vacuum, and $\varepsilon \approx 80$ is the relative dielectric permittivity of water, i.e. by 5 - 6 orders of magnitude shorter than the characteristic time of the excitation and relaxation of the action potential in the axons. The size of the non-quasi-neutral region near the membrane of axons is determined by the Debye length $\lambda_D = \sqrt{\varepsilon\varepsilon_0 k_B T / \left( \sum\limits_{j=1}^{J} n_j q_j^2 \right)}$. Here $k_B$ it the Boltzmann constant, $T$ is the temperature, $n_j$ are the densities of the ions with the charges $q_j$. In the case of saline, we can assume that all the ions in the liquid are singly charged and their total density is of the order $n_j \approx 2 \cdot 10^{26} \; \mathrm{m}^{-3}$ [25]. For $T \approx 300\,\mathrm{K}$, $\lambda_D \approx 0.5\,\mathrm{nm}$ and this value is much smaller than the typical radius of the axon $R_0 \sim 3-10\,\mu\mathrm{m}$. Therefore, the violation of quasi-neutrality cannot be taken into account and the potential distribution in the vicinity of a neuron can be found from the equation of continuity of current

$$\nabla \cdot \vec{J} = 0 \, , \quad \vec{J} = \sigma\vec{E} \, . \tag{4}$$

Since the conductivity $\sigma$ of the electrolyte is constant and $\vec{E} = -\nabla\varphi$, the problem of determining the potential is reduced to the Laplace equation:

$$\Delta\varphi = 0 \, , \tag{5}$$

with Neumann boundary conditions on the surface of the membrane:

$$\frac{\partial\varphi}{\partial r}\bigg|_{r=R_0} = \frac{J_m}{\sigma} \, , \tag{6}$$

where $J_m$ is the total radial current through the excited membrane of the axon, and the Dirichlet condition, far away from the membrane surface:

$$\varphi_{r \to \infty} = 0 \, . \tag{7}$$

The potential distribution in saline outside the excited neuron we will seek as in [10,11], in which the magnetic field accompanying the action potential was calculated. However, in our case, in contrast to [10,11], it was assumed be given the currents through the non-myelined parts of the axon (nodes of Ranvier and the initial segment): the current $J_m$ is equal to the sum of the ion, $J_{ion}$, and the capacitive, $J_{capas} = C_m \cdot \partial U / \partial t$ currents, where $C_m \approx 2 \cdot 10^{-6} [\text{F/cm}^2]$ is the capacity of the membrane per unit area; $U = V_{AP} - V_R$ is the voltage between the inner and outer surfaces of the membrane during the passage of the action potential. But at the myelined sections: $J_m = J_{capas}$, because the ion current can be neglected there, as compared with the capacitive current. Assuming, as in [10,11], that the membrane of the axon is a thin-layer cylindrical capacitor, it follows from (5) that outside the axon ($r \geq R_0$) the potential is:

$$\varphi(r,z,t) = \frac{1}{2\pi} \int_{-\infty}^{+\infty} \varphi_{e,k}(t) \cdot \frac{K_0(|k| \cdot r)}{K_0(|k| \cdot R_0)} e^{-ik \cdot z} dk \qquad (8)$$

Here $\varphi_{e,k} = \int_{-\infty}^{\infty} \varphi_e(z,t) \cdot e^{ikz} dz$ is the Fourier transform of the potential on the outer surface of the membrane, $K_0(|k| \cdot r)$, are modified Bessel functions of the second kind, of zero and first order, respectively. Since the currents and potential outside the membrane are related by (6), then, respectively:

$$J_r(r,z,t) = \frac{1}{2\pi} \int_{-\infty}^{+\infty} \frac{K_1(|k| \cdot r)}{K_1(|k| \cdot R_0)} \cdot J_{m,k}(t) e^{-ik \cdot z} dk \qquad (9)$$

$$J_z(r,z,t) = \frac{i}{2\pi} \int_{-\infty}^{+\infty} \frac{k}{|k|} \frac{K_0(|k| \cdot r)}{K_1(|k| \cdot R_0)} \cdot J_{m,k}(t) e^{-ik \cdot z} dk \qquad (10)$$

Here $J_r(r,z,t)$ и $J_z(r,z,t)$ the radial and longitudinal components of the current $\bar{J}$ outside the the axon, $J_{m,k}(t) = \int_{-\infty}^{\infty} J_m(z,t) e^{ikz} dz$ is the Fourier transform of the radial current through the membrane.

## IV. The condition for the spike initiation in the initial segment by a spike in neighboring neuron

Without loss of generality, we assume that the radial current through the membrane of the initial segment of the length $L_{IS}$ has a Gaussian distribution: $J_m(t,z) = J_0(t) \exp(-4z^2 / L_{IS}^2)$. Accordingly, the Fourier component of the transversal current $J_{m,k} = \sqrt{\pi} L_{IS} / 2 \cdot J_0(t) \exp(-k^2 \cdot L_{IS}^2 / 16)$. In this case:

$$J_r(r,z,t) = \frac{2L_{IS} J_0(t)}{\sqrt{\pi}} \int_{0}^{+\infty} \frac{K_1(k \cdot r)}{K_1(k \cdot R_0)} \exp(-k^2 \cdot L_{IS}^2 / 16) \cdot \cos(k \cdot z) dk \qquad (11)$$

$$J_z(r,z,t) = -\frac{2L_{IS}J_0(t)}{\sqrt{\pi}}\int\limits_0^{+\infty}\frac{K_1(k\cdot r)}{K_1(k\cdot R_0)}\exp\left(-k^2\cdot L_{IS}^2/16\right)\cdot\sin(k\cdot z)dk \qquad (12)$$

Since the initial segment and the nodes of Ranvier are the axon areas not covered by the myelin sheath, for the approximate description of the action potential therein we will use the Hodgkin-Huxley equations for the non-myelined squid axon [26]. For definiteness, assume that the surface densities of ion channels in the initial segment and the nodes of Ranvier are the same as in [26]. Figure 6 shows the time dependence of the radial current through a membrane obtained on the basis the equations of Hodgkin and Huxley [26,19]. The current $J_0$ in equations (6) and (7) corresponds to the total current, which is equal to the sum of the capacitive and the ions currents. Figure 7 shows the radial distribution of the currents at $z = 0$, the radius of the membrane $R_0 = 4.5\,\mu\text{m}$. Figure 8 shows the contour lines of the currents amplitudes $J_e = \sqrt{J_r^2(r,z)+J_z^2(r,z)}$ for $L_{IS} = 30$ and $70\,\mu\text{m}$.

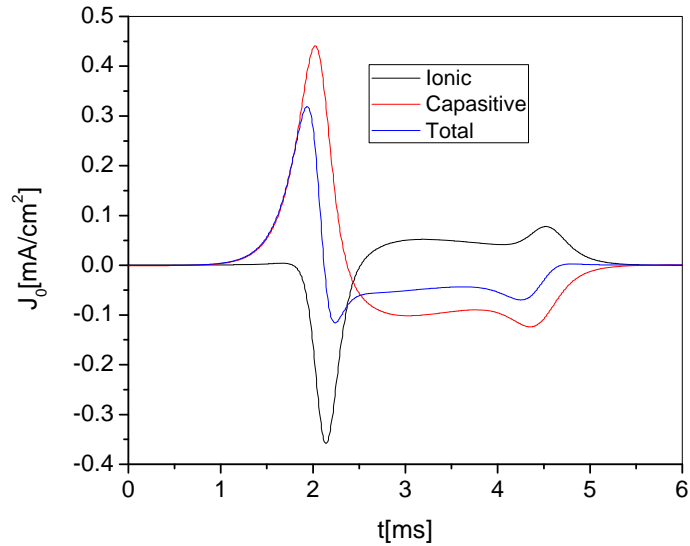

**Fig.6.** The time dependencies of the radial currents in the membrane obtained on the basis of Hodgkin and Huxley equations.

The potential difference on the inactive membrane (see Figure 1) without the draining of charge is determined by the charge on its surface. Knowing the dependence of the charging current on time and its distribution in space, it is easy to calculate the potential difference across the membrane of the inactive axon initial, depending on its position:

$$\Delta\varphi_m(z,r) = \frac{1}{c_m}\int\limits_0^{\infty}\left(\vec{J}\cdot\vec{n}\right)dt\;. \qquad (13)$$

Figure 9 shows the areas of synchronization of the action potentials for the axons in the case $L_{IS} = 70\,\mu m$ and assuming average initiation potentials of $U_s = 5$ , 10 and 20 mV and the initiation time interval of $\delta t = 0.1\,ms$ .

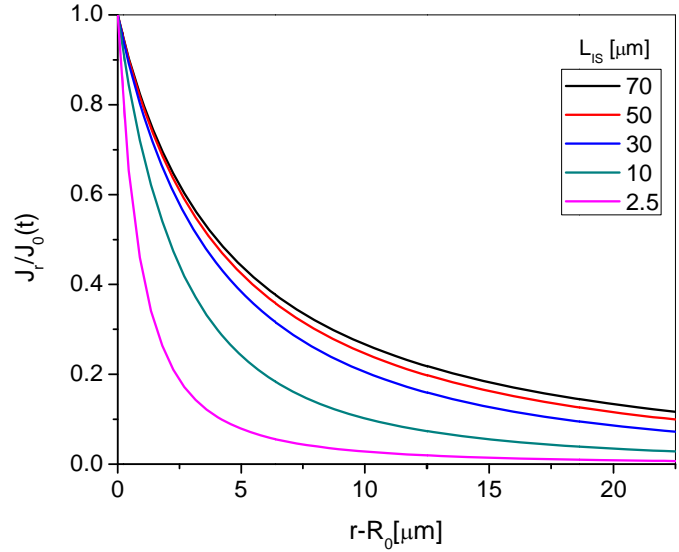

**Fig.7.** The radial dependencies of the current in the point $z = 0$ for different length of the initial segment. $R_0 = 4.5\,\mu m$ .

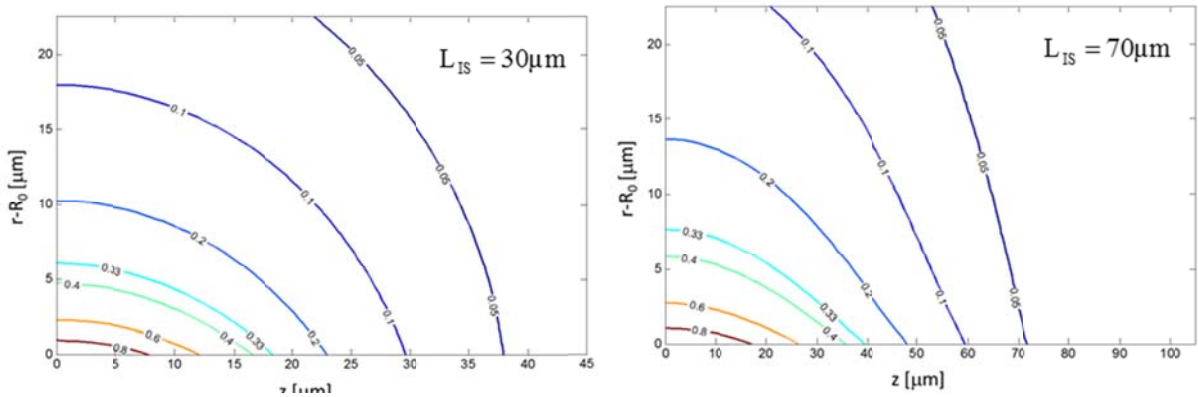

**Fig.8.** The instantaneous normalized amplitude distributions of the current density $J_e = \sqrt{J_r^2(r,z) + J_z^2(r,z)}$ in the vicinity of the initial segment of lengths: (a): $L_{is} = 30$ and (b): 70 $\mu m$.

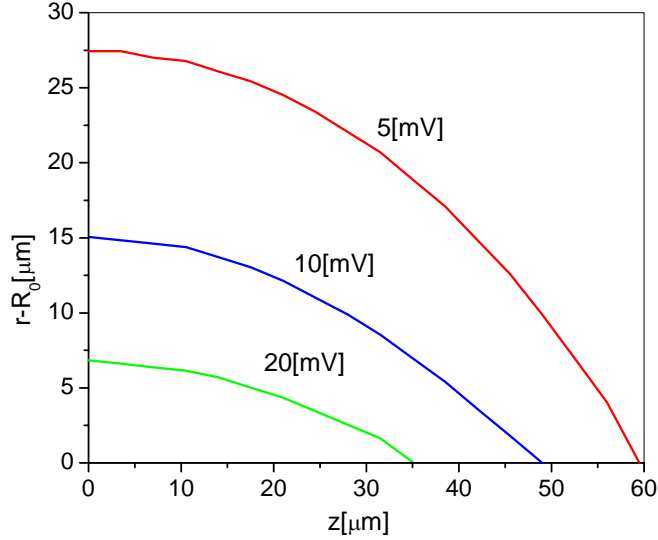

**Fig.9**. Areas of neurons synchronization at $L_{IS} = 70\,\mu\text{m}$, the assumed initiation time $\delta t = 0.1\,\text{ms}$ and at assumed average initiation voltage $U_s = 5$, 10 and 20 mV, correspondingly.

## V. The condition of the action potential initiation in the myelin fiber by the action potential in the neighboring neuron.

As was noted above, the radial current through the myelin section of the axon membrane is capacitive $J_m = C_m \cdot \partial U / \partial t$ (the contribution of the ion current can be neglected because the length of the node of Ranvier is almost two orders less than the length of the myelin section). Fig. 10 shows the spatial distribution of the action potential $U(z,t) = V_{AP}(z,t) - V_R$ along the myelin fiber [15,18] and the corresponding capacitive current $J_m = C_m \cdot \partial U / \partial t$ through the membrane. The dashed curve shows the current approximation by a Gaussian distribution. Under the assumption of a Gaussian distribution of the membrane currents: $J_m(t,z) = J_0 \exp\left(-\left(z - v \cdot t\right)^2 / L^2\right)$, where $J_0$ is the maximal current through the membrane, $v$ is the velocity of the action potential along the fiber, we obtain expressions for the currents in saline outside the axon:

$$J_r(r,z) = \frac{J_0 L}{\sqrt{\pi}} \int_0^{+\infty} \frac{K_1(k \cdot r)}{K_1(k \cdot R_0)} \exp\left(-k^2 L^2 / 4\right) \cos\left(k \cdot (z - vt)\right) dk \qquad (14)$$

$$J_z(r,z) = \frac{J_0 L}{\sqrt{\pi}} \int_0^{+\infty} \frac{K_0(k \cdot r)}{K_1(k \cdot R_0)} \exp\left(-k^2 L^2 / 4^2\right) \sin\left(k \cdot (z - vt)\right) dk \qquad (15)$$

For these parameters of the myelin fiber and assuming the same conditions as in [17,20], $v = 18.4$ m/s, $L \approx 2.2$ mm. Figure 11 shows the radial distribution of the current normalized to the maximum current at the point $z = 0$, for the case of a Gaussian distribution along z-axis (curve 2 in Fig. 10b). Figure 11 shows examples of the longitudinal distribution of the charging current amplitude corresponding to the current $J_e = \sqrt{J_r^2 + J_z^2}$ at different distances from the surface of the membrane of the active axon. Figure 12 shows the dependence of the potential difference on time for the membrane of the inactive axon at different distances from the membrane of the active axon.

It follows from the results shown in Fig. 9 and 12 that at $U_s = 5\,\text{mV}$ the radius of synchronization is $R_s = 55\,\mu\text{m}$; at $U_s = 10\,\text{mV}$, $R_s = 24.3\mu\text{m}$, and, at $U_s = 20\,\text{mV}$, $R_s = 9.7\mu\text{m}$.

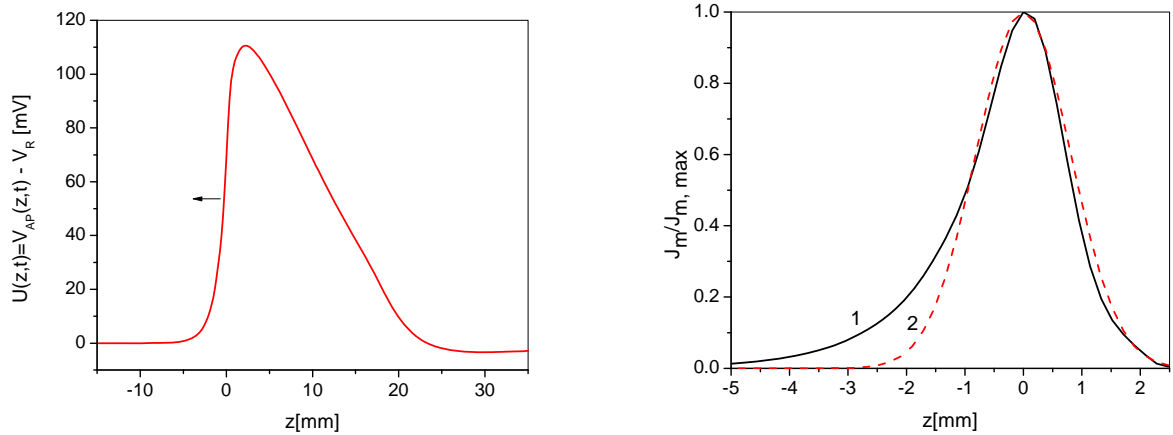

**Fig. 10.** (a) The dependence of the action potential on z (propagating from left to right); (b) the distribution displacement current (curve 1) and the approximation of this current by a Gaussian distribution (curve 2).

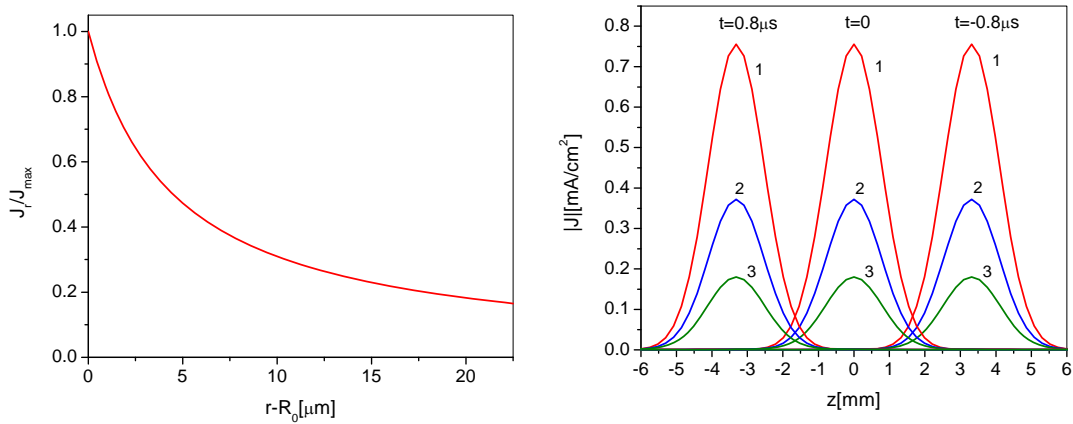

**Fig. 11.** (a) The radial dependence of the current at the point $z = 0$, $R_0 = 4.5\,\mu\text{m}$ for the case of a Gaussian distribution along z-axis (curve 2 in Fig.5b); (b) Examples of the longitudinal distribution

of the charging current $J_e$ at the time moments $t = -0.8\,\mu s$, $0$, and $0.8\,\mu s$. Curve 1 corresponds to the current at a distance $r - R_0 = 9.7\,\mu m$ from the surface of the membrane; $2 - r - R_0 = 24.3\mu m$; $3 - r - R_0 = 55\mu m$

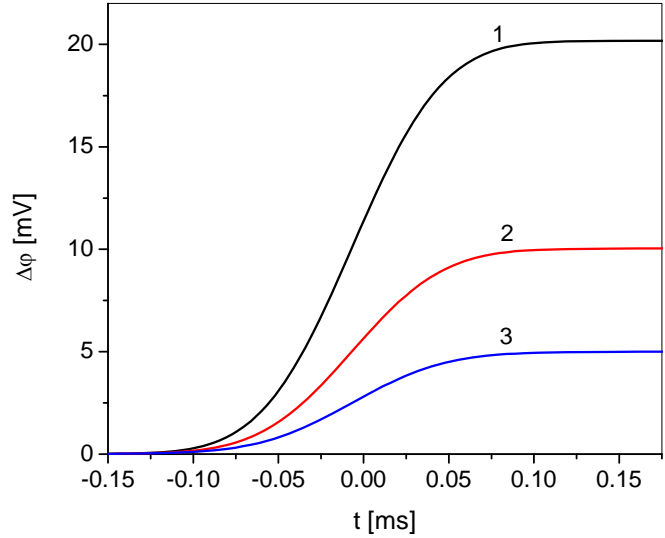

**Fig. 12.** The dependence on time of the potential difference at the membrane of the inactive axon for different distances from the active axon membrane. Curve 1 corresponds to the current at a distance $r - R_0 = 9.7\,\mu m$ from the surface of the membrane; $2 - r - R_0 = 24.3\mu m$; $3 - r - R_0 = 55\mu m$

## VI. A possible experimental condition for initiation of the action potential in the squid axon by the action potential in the neighboring squid axon.

Within the body of the giant squid, axons are located at a distance of a few tens of centimeters, so there cannot be a correlation of the action potentials between these axons. However, the axons of squid are a traditional and convenient object of experimental studies. We estimate the size of the correlation area for the action potentials in the squid axons, which is possible to observe in a laboratory experiment. Fig. 13 shows the z-distribution of the total current through the membrane of the squid axon and its approximation by a Gaussian distribution. For propagating action potentials, $J_{total} = J_{total}(z + vt)$, where $v \sim 15\,\text{m/s}$ is the velocity of propagation of the action potential in the axon of a squid. Since we are interested only in the stage of "charging" of the axon's membrane, which is located in the field of currents initiated by the action potential propagation in adjacent axons, it is enough to restrict ourselves to the Gaussian approximation, shown in Figure 13. Figure 14 shows the radial distribution of the current $J_r$ normalized to the maximum current at the point $z = 0$ for the case of a Gaussian distribution along z-axis (curve 2 in Figure 6b). Figure 14 also shows examples of the longitudinal distribution of the charging current. The dependence of the potential difference on the membrane of the inactive axon time at different distances from the active axon membrane is shown in Figure 15. In this case, the radius of synchronization $R_s = 1.96\,\mathrm{mm}$ at $U_s = 5\,\mathrm{mV}$; $R_s = 0.85\,\mathrm{mm}$ at $U_s = 10\,\mathrm{mV}$; and $R_s = 0.27\,\mathrm{mm}$ at $U_s = 20\,\mathrm{mV}$.

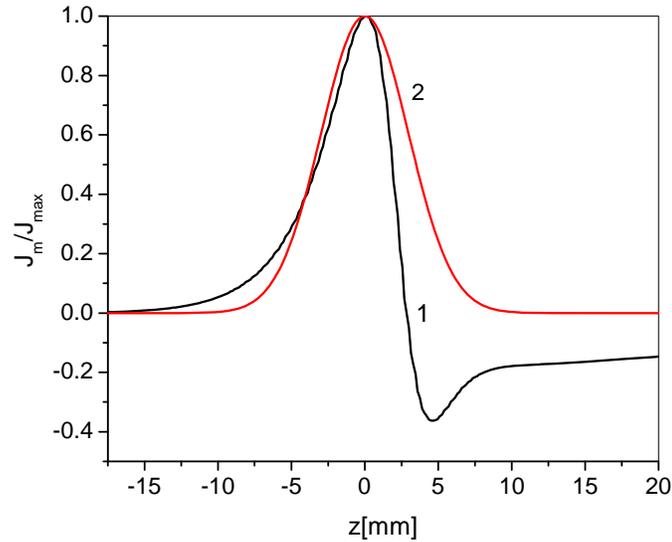

**Fig. 13.** z-distribution of the total current through the membrane of the axon of squid (curve 1) and its approximation by a Gaussian distribution (curve 2) for propagating action potential, v=15 m/s.

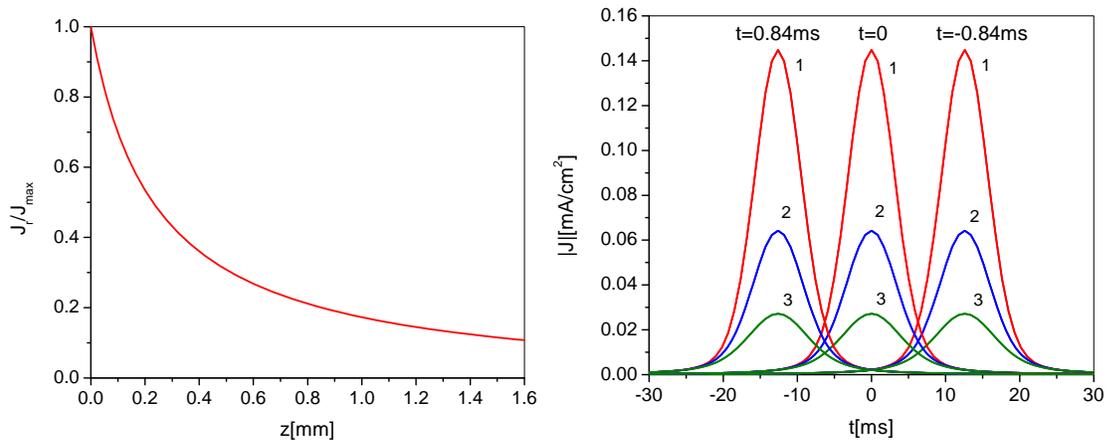

**Fig. 14.** (a) The radial dependence of the current at the point $z = 0$, $R_0 = 0.24\,\mathrm{mm}$ for the case of a Gaussian distribution along z (curve 2 in Figure 13). (b) examples of the longitudinal distribution of the charging current $J_e$. Curve 1 corresponds to the current at a distance $r - R_0 = 0.27\,\mathrm{mm}$ from the surface of the membrane, $2 - r - R_0 = 0.84\,\mathrm{mm}$, $3 - r - R_0 = 1.96\,\mathrm{mm}$.

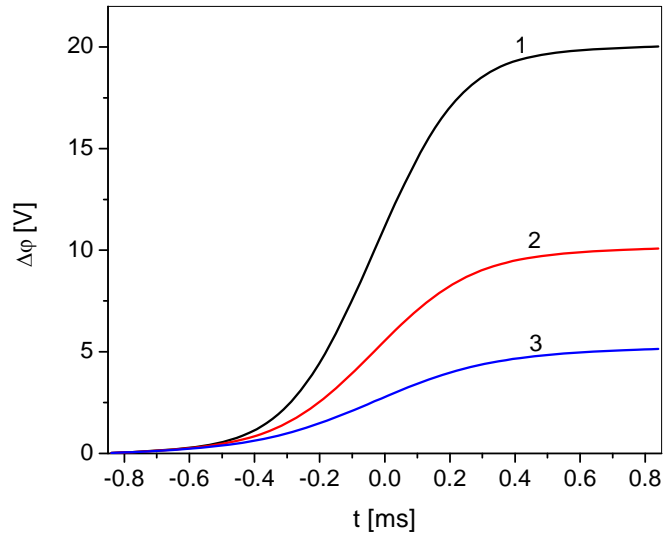

**Fig. 15.** The dependence on time of the potential difference at the membrane of the inactive axon for different distances from the active axon membrane. Curve 1 corresponds to the current at a distance $r - R_0 = 0.27$ mm from the surface of the membrane; **2** − $r - R_0 = 0.84$ mm, **3** − $r - R_0 = 1.96$ mm.

**Conclusions**

1. Simple analytical expressions were obtained for the action potential propagation along the myelined section of an axon and the maximal length of the myelined coating, which is necessary for the saltatory conduction, are obtained.
2. A synchronization mechanism for action potentials in the nearby neurons is proposed.
3. It is shown that in the vicinity of the initial segment of the excited neuron, there is a noticeable area in which currents originating in interstitial saline are sufficient for the charging of excitable membranes of initial segments or axons of neighboring neurons up to initiation of the action potential. This distance can be several times greater than the radius of the fiber, which corresponds to the experiments [1].
4. The conditions for the synchronization of the action potentials in squid axons are obtained. These results can be used for laboratory testing of the mechanism proposed in this paper.

**References**

1. C.A. Anastassiou, R. Perin, H. Markram**,** C. Koch**,** Ephaptic coupling of cortical neurons, Nature neuroscience. 14 (2) 217 (2011)
2. B. Katz, O,H. Schmitt, Electric Interaction Between Two Adjacent Nerve Fibers, *J Physiol* 97 (4): 471–488 (1940)

3. Arvanitaki A, Effects evoked in an axon by the activity of a contiguous one, J Neurophysiol 5:89–108, 1942.

4. A. Scott, Neuroscience. A Mathematical Primer, Springer (2002)

5. H. Bokil, N. Laaris, K. Blinder, M. Ennis, and A. Keller, Ephaptic Interactions in the Mammalian Olfactory System, J. Neurosci., , Vol. 21 RC173 (2001)

6. R. Costalat, G. Chauvet, Basic properties of electrical field coupling between neurons: an analytical approach, Journal of Integrative Neuroscience, 7, 225–247 (2008)

7. D. Debanne, E. Campanac, A. Bialowas, E. Carlier, G. Alcaraz, Axon Physiology, Physiol Rev 91, 555–602 (2011)

8. L.J. Colwell, M.P. Brenner, Action Potential Initiation in the Hodgkin-Huxley Model. PLoS Comput. Biol., **5** (1), e1000265: 1-7 (2009)

9. J. Platkiewicz, R. A. Brette Threshold Equation for Action Potential Initiation. PLoS Comput. Biol. **6**, e1000850: 1-16 (2010)

10. K.R. Swinney J.P. Wikswo, Jr. "A calculation of the magnetic field of a nerve action potential", Biophysical Journal,  V 32, 719-732 (1980).

11. B.J. Roth, J.P. Wikswo, Jr. The magnetic field of a single axon, Biophysical Journal,  V 48, 93-109 (1985).

12. R.C. Barr, R. Plonsey Electrophysiological interaction through interstitial space between adjacent unmyelinated parallel fibers. Biophys J., 61:1164–1175 (1992)

13. J.P. Segundo, What can neurons do to serve as integrating devices. J Theor Neurobiol 5, 1–59 (1986)

14. D.W. Esplin, Independence of conduction velocity among myelinated fibers in cat nerve. J Neurophysiol 25:805–811(1962)

15. I. Tasaki, Electro-saltatory transmission of nerve impulse and effect of narcosis upon nerve fiber". Amer. J. Physiol. 127: 211–27 (1939)

16. A. F. Huxley and R. Stampfli, Evidence for saltatory conduction in peripheral myelinated nerve fibers, J. Physiol. 108, 315 (1949)

17. L. Goldman, J. S. Albus, Biophys. J. Computation of impulse conduction in myelinated fibers, 8, 596 (1968)

18. D.K. Hartline, D.R. Colman, Rapid conduction and the evolution of giant axons and myelinated fibers, Curr. Biol. 17, R29 (2007)

19. B. Frankenhaeuser, A. F. Huxley, The action potential in the myelinated nerve fibre of xenopus laevis as computed on the basis of voltage clamp data, J. Physiol. (London) 171, 302 (1964)

20. M. N. Shneider, A. A. Voronin, A. M. Zheltikov, Modeling the action-potential-sensitive nonlinear-optical response of myelinated nerve fibers and short-term memory, J. Appl. Phys. 110, 094702 (2011)

21. H. S. Carslaw, J. C. Jaeger, Conduction of Heat in Solids (Oxford Science Publications, 1986)

22. A.D. Polyanin, Handbook of Linear Partial Differential Equations for Engineers and Scientists (Boca Raton–London: Chapman & Hall/CRC Press, 2002)

23. R. Glaser, Biophysik (Berlin, Heidelberg, New York, Springer-Verlag, 1996)

24. Yu.P. Raizer, Gas Discharge Physics (Berlin, Heidelberg, New York, Springer-Verlag, 1991)

25. I.P. Herman, Physics of the Human Body (Springer, Berlin Heidelberg, 2007)

26. A.L. Hodgkin, A.F. Huxley, A quantitative description of membrane current and its application to conduction and excitation in nerve, J. Physiol. 117, 500-544 (I952)